# Rational design principles of quantum anomalous Hall effect from superlattice-like magnetic topological insulators

Hongyi Sun[1], Bowen Xia[1], Zhongjia Chen[1,2], Yingjie Zhang[1], Pengfei Liu[1], Qiushi Yao[1], Hong Tang[1], Yujun Zhao[2], Hu Xu[1] and Qihang Liu[1,3,*]

[1]*Shenzhen Institute for Quantum Science and Technology and Department of Physics, Southern University of Science and Technology (SUSTech), Shenzhen, 518055, China*
[2]*Department of Physics, South China University of Technology, Guangzhou 510640, China*
[3]*Center for Quantum Computing, Peng Cheng Laboratory, Shenzhen 518055, China*
H.S., B.X. and Z.C. contributed equally to this work.
[*]Email: liuqh@sustech.edu.cn

**Abstract**

As one of paradigmatic phenomena in condensed matter physics, the quantum anomalous Hall effect (QAHE) in stoichiometric Chern insulators has drawn great interest for years. By using model Hamiltonian analysis and first-principle calculations, we establish a topological phase diagram and map on it with different two-dimensional configurations, which is taken from the recently-grown magnetic topological insulators $MnBi_4Te_7$ and $MnBi_6Te_{10}$ with superlattice-like stacking patterns. These configurations manifest various topological phases, including quantum spin Hall effect with and without time-reversal symmetry, as well as QAHE. We then provide design principles to trigger QAHE by tuning experimentally accessible knobs, such as slab thickness and magnetization. Our work reveals that superlattice-like magnetic topological insulators with tunable exchange interaction serve as an ideal platform to realize the long-sought QAHE in pristine compounds, paving a new avenue within the area of topological materials.

The marriage between magnetism and the topology of electronic structures in condensed matter systems provides fruitful ground for the exploration of exotic quantum phenomena [1-3]. Among them, the quantum anomalous Hall effect (QAHE), induced by spontaneous magnetization without an extrinsic magnetic field, have been sought for years because its dissipationless chiral edge state holds the potential to realize the next-generation electronic devices with ultralow power cost [4,5]. The QAHE was firstly observed in Cr-doped topological insulator (TI) thin films, but only at low temperature around 100 mK [6]. Therefore, many efforts, especially theoretical proposals [7-14], have been put into 2D stoichiometric magnetic insulators with a nontrivial Chern number (i.e., Chern insulators) in order to realize QAHE that survives at a higher temperature, yet without solid experimental confirmation.

Recent breakthrough in such direction has been made thanks to the growth technology of 2D ferromagnetic (FM) semiconductors [15,16] and more importantly, intrinsic 3D magnetic TI $MnBi_2Te_4$, a magnetic version in analogy to time-reversal ($T$) preserved TI $Bi_2Te_3$ stacked by Van der Waals (VdW) connected sublayers. It is predicted and soon experimentally verified that $MnBi_2Te_4$ in its magnetic ground state is a $Z_2$ antiferromagnetic (AFM) TI protected by a combined symmetry of $T$ and a fractional translation operation [17-21]. Later, the quantized Hall conductance is observed in the 2D limit of $MnBi_2Te_4$ down to 5-6 septuple layers [22,23]. However, although 2D $MnBi_2Te_4$ multilayers with uncompensated AFM configurations are also predicted to be Chern insulators [24], a magnetic field about 10 Tesla is required to fully align the spins to a FM state [22,23]. Strictly speaking, the QAHE happens without magnetic field because it originates from the topological protected chiral edge states of a Chern insulator [5]. The external magnetic field brings in two types of ambiguity: one is the contribution from the quantized Landau levels, i.e., quantum Hall effect [25]; and the other is the possible topological transition from normal insulator (NI) to Chern insulator induced by the Zeeman exchange field [26]. Hence, to obtain an ideal QAHE system with, if necessary, a small enough magnetic field, a better understanding of the route to achieve the QAHE with experimentally tunable "knobs", as well as realistic material candidates are highly desirable.

In this Letter, we investigate the topological phase transitions of 2D magnetic systems and how to rationally design quantum anomalous Hall insulators. Starting from an effective model of 3D $T$-preserved TI (such as Bi$_2$Te$_3$) under spontaneous exchange field, we present solutions for 2D thin films and thus a phase diagram containing various topological states including $T$-preserved quantum spin Hall effect (QSHE) [27,28], $T$-broken QSHE [29], QAHE phase. For material realization, we are inspired by the recent growth of MnBi$_4$Te$_7$ and MnBi$_6$Te$_{10}$ single crystals [30,31], which are in essence 1:1 and 1:2 superlattice composed by MnBi$_2$Te$_4$ septuple layer (denoted as "A" hereafter) and Bi$_2$Te$_3$ quintuple layer (denoted as "B" hereafter), respectively. Thanks to the diversity of 2D configurations of MnBi$_4$Te$_7$ and MnBi$_6$Te$_{10}$ compared with MnBi$_2$Te$_4$, we find by using first-principle calculations that the combinations of A and B building blocks are able to realize all the topological phases in our phase diagram. For example, ABA in its FM state exhibits QAHE, while AB and BAB exhibits $T$-broken QSHE. Since the Bi$_2$Te$_3$ buffer layer effectively decreases the AFM coupling between two neighboring A layers, the magnetic field required to trigger the QAHE is orders of magnitude weaker than that for MnBi$_2$Te$_4$ thin films. Furthermore, we illustrate the possibilities to manipulate the topological phase transitions in the phase diagram either horizontally (by stacking) or vertically (by exchange field) through band gap engineering.

*Topological phase diagram from 2D effective model*

Staring from the 4 × 4 model Hamiltonian of $T$-preserved 3D TI for Bi$_2$Se$_3$ and Bi$_2$Te$_3$ [32], we scale it down to the 2D limit by replacing $k_z$ with the $\partial_z$ operator [33], and then add a uniform exchange field to it. The resultant model Hamiltonian is thus written as:

$$H_{2D} = \epsilon(\mathbf{k}) + \begin{pmatrix} M(\mathbf{k}) + gM & B_0\partial_z & 0 & A_0 k_- \\ B_0\partial_z & -M(\mathbf{k}) + gM & A_0 k_- & 0 \\ 0 & A_0 k_+ & M(\mathbf{k}) - gM & -B_0\partial_z \\ A_0 k_+ & 0 & -B_0\partial_z & -M(\mathbf{k}) - gM \end{pmatrix}, \quad (1)$$

where $\epsilon(\mathbf{k}) = C_0 - C_1\partial_z^2 + C_2(k_x^2 + k_y^2)$, $M(\mathbf{k}) = M_0 - M_1\partial_z^2 + M_2(k_x^2 + k_y^2)$. The Zeeman effect is characterized by exchange field $M$ along the $z$ axis and the effective $g$ factor. This Hamiltonian Eq. (1) is invariant under $T$ as well as inversion symmetry. The extrinsic structural inversion asymmetry (SIA) induced by substrate effect is not included here, but the SIA effect from the specific material candidates will be discussed later.

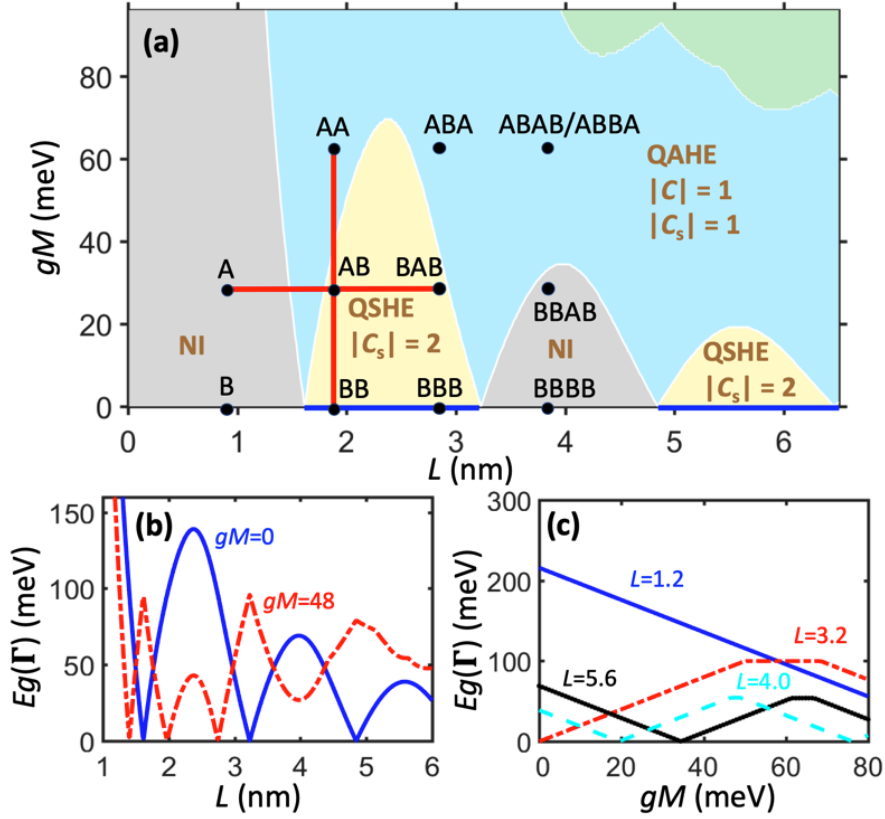

Fig. 1: (a) Topological phase diagram of the 2D limit of 3D TI under an exchange field, in terms of the film thickness $L$ and the magnetization $gM$. Various phases including NI (grey area), $T$-preserved QSHE (blue line) $T$-broken QSHE (yellow area), and QAHE (cyan area) are shown. The DFT calculated 2D configurations composed by $MnBi_2Te_4$ (A) and $Bi_2Te_3$ (B) building blocks are also mapped on the phase diagram (also see Table I and Fig. 2). (b,c) Energy gap at the $\Gamma$ point $E_g(\Gamma)$ as a function of (b) the film thickness $L$ and (c) the magnetization $gM$.

We then numerically solve the continuous model Hamiltonian Eq. (1) and obtain the topological phase diagram as functions of the thickness of the film $L$ and the magnetization $gM$, as shown in Fig. 1(a). The boundary separating different phases is indeed the trajectory of zero gap at the $\Gamma$ point, indicating that the topological phase transition must be accompanied with gap closing and reopening. We next determine the topological properties of each area. Taking a fixed thickness $L$ and only four bands around the Fermi level, Eq. (1) is reduced to a block-diagonal Hamiltonian with two spin channels decoupled with each other, in an equivalent form discussed in Ref. [26,34]:

$$H_{eff}(\mathbf{k}) = (E_0 - Dk^2)\tau_0\sigma_0 + (\Delta - Bk^2)\tau_z\sigma_z - \gamma\tau_0(k_x\sigma_y - k_y\sigma_x) + gM\tau_0\sigma_z, \quad (2)$$

where $\boldsymbol{\sigma}$ denotes the Pauli matrices for spin and $\boldsymbol{\tau}$ for bonding and antibonding states of the two surfaces. The other parameters of Eq. (2) as functions of film thickness $L$ are numerically solved from Eq. (1), as shown in Supplementary Section I. Thus, the Chern numbers of the two spin channels can be analytically solved as $C_+ = \frac{1}{2}[sgn(\Delta + gM) + sgn(B)]$ and $C_- = \frac{1}{2}[sgn(-\Delta + gM) - sgn(B)]$, respectively. The Chern number of the whole system is $C = C_+ + C_- = \frac{1}{2}[sgn(\Delta + gM) + sgn(-\Delta + gM)]$, while the spin Chern number (for $\tau_z$) $C_s = C_+ - C_- = \frac{1}{2}[sgn(\Delta + gM) - sgn(-\Delta + gM)] + sgn(B)$. The phase diagram is then established according to the topological invariants.

We note that the $T$ only preserves when $M = 0$, for which the topological nature of the thin film exhibits here an oscillatory behavior between NI and $T$-preserved QSHE [35,36]. With a finite magnetization $|gM| < \Delta$, there is no gap closing between $T$-preserved QSH and $T$-broken QSH, indicating that they are equivalent topological phases characterized by the same spin Chern numbers. Several observations can be drawn from Fig. 1. Firstly, a large enough magnetization $M$ can drive either NI or QSH insulator to the QAHE territory, in consistent with Ref. [26]. Therefore, even the effect of Landau level can be excluded, the magnetic-field induced quantized Hall conductance holds the ambiguity that the magnetic field can trigger the QAHE by such phase transition. Secondly, the phase transition between NI and $T$-broken QSHE is inevitably accompanied with a QAHE region. This can be understood by the evolution of band inversion for different spin channels. From NI to $T$-broken QSHE, in which the two spin channels are nonequivalent, the sequent band inversion by tuning the order parameter $L$ naturally leads to a QAHE region with the inverted band order for only one spin channel. Thirdly, the critical points along the $L$ axis are triple-phase points among NI, QSHE, and QAHE. These points are the only connection between NI and QSHE, while a tiny magnetization drives the phase to the QAHE region [26,34]. Finally, for a thick slab and strong magnetization, there is another trajectory for $E_g(\Gamma) = 0$, depicting an island with distinguished topological phase. It is because the subbands of the quantum well system with higher energy move to the Fermi level by the exchange field, and thus induce another band inversion. Such area is not our focus here because it is challenging to realize experimentally.

The band gaps at the Γ point $E_g(\Gamma)$ as functions of the film thickness $L$ and the magnetization $gM$ are shown in Fig. 1(b) and 1(c), respectively. For $M = 0$, $E_g(\Gamma)$ oscillates with a gradually decreased amplitude approaching zero in terms of $L$ [37]. The period of the oscillation is $\sim \pi\sqrt{M_1/|M_0|}$ (1.6 nm). With nonzero magnetization, $E_g(\Gamma)$ remains open after a certain thickness, and the number of zero-gap nodes is determined by the number of NI or QSHE regions passed by in the phase diagram. For example, the evolution of $E_g(\Gamma)$ at $gM = 48$ meV indicates a NI-QAHE-QSHE-QAHE phase transition. On the other hand, with a fixed $L$, $E_g(\Gamma)$ generally closes only once by increasing $M$ within a moderate range, corresponding to the transition between NI (or QSHE) and QAHE.

Table I: Properties of considered 2D configurations, including the number of layers, stacking pattern of building blocks, parent 3D compounds, energy difference $\Delta E$ between FM and AFM phase, energy gap at the Γ point $E_g(\Gamma)$, Chern number ($C$), and spin Chern number ($C_s$).

| # of layers | Stacking | Parent compound | $\Delta E$ (meV/Mn) | $E_g(\Gamma)$ (meV) | $C$ | $C_s$ |
|---|---|---|---|---|---|---|
| 2 | AB | Both | 0 | 158.9 | 0 | 2 |
| 3 | ABA | MnBi$_4$Te$_7$ | 0.17 | 37.4 | 1 | 1 |
|   | BAB | MnBi$_4$Te$_7$ | 0 | 122.9 | 0 | 2 |
|   | ABB | MnBi$_6$Te$_{10}$ | 0 | 36.7 | 0 | 2 |
| 4 | ABAB | MnBi$_4$Te$_7$ | 0.15 | 12.8 | 1 | 1 |
|   | ABBA | MnBi$_6$Te$_{10}$ | 0.03 | 77.1 | 1 | 1 |
|   | BBAB | MnBi$_6$Te$_{10}$ | 0 | 16.0 | 0 | 0 |

*Realizing various topological phases by MnBi$_2$Te$_4$ (A) and Bi$_2$Te$_3$ (B) building blocks*

Our material realization of different topological phases is inspired by the recent growth of MnBi$_4$Te$_7$ and MnBi$_6$Te$_{10}$ single crystals [30,31] formed as superlattices with AB and ABB stacking pattern, where A and B are building blocks of 3D AFM TI MnBi$_2$Te$_4$ and 3D nonmagnetic TI Bi$_2$Te$_3$, respectively. The calculations are performed by density functional theory (DFT) with the presence of spin-orbit coupling (SOC), with the details provided in Supplementary Section II. As shown in Fig. 2(a) and Table I, we only consider the 2D configurations that are certain fragments of the 3D parent $MnBi_4Te_7$ or $MnBi_6Te_{10}$ for the ease of exfoliation. The exfoliation energy of these slabs is aound 22 meV/Å$^2$ (see Supplementary Section III), comparable to that of graphite (18 meV/Å$^2$). For the configurations with only one A layer, the FM phase is the ground state, while for those with two A layers, the FM phase can be stablized by a tiny magnetic field because its total energy is only ~ 0.1 meV (or less) higher than the ground state (see Table I), i.e., AFM coupling between neighbouring A layers. Due to the role of B layer as spacer, the out-of-plane saturation field of bulk $MnBi_4Te_7$ was measured as low as 0.22 T, 40 times lower than that of $MnBi_2Te_4$ [30]. With a brief note that the AFM phases with even number of A layers are candidates of axion insulators with zero-plateau QAHE [18,24], we next focus on FM phases for all configurations. The diversity of of film thickness and magnetization strength helps us to establish the parametrized connection between different topological phases based on the phase diagram, and thus the design principle of the QAHE.

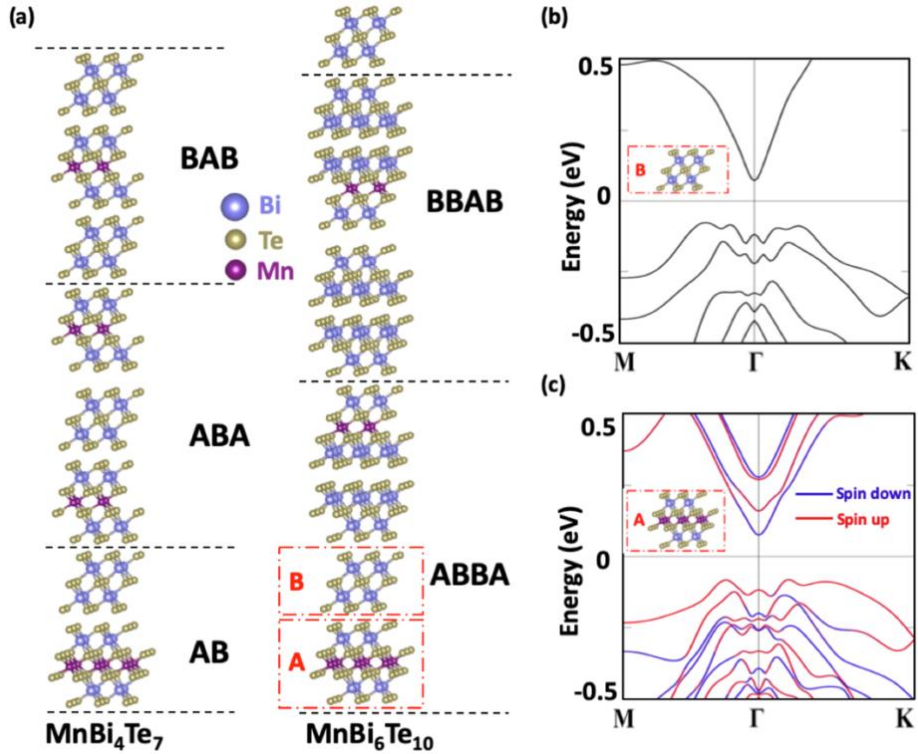

Fig. 2: (a) The crystal structures of $MnBi_4Te_7$ and $MnBi_6Te_{10}$ with their different fragments (marked by black dashed lines) that can be exfoliated as 2D structures. The red dashed

boxes denote the A (MnBi$_2$Te$_4$ septuple layer) and B (Bi$_2$Te$_3$ quintuple layer) building blocks. (b) Band structure of the B layer. (c) Band structure of the A layer with the projection of spin operator $\sigma_z$. Red and blue denote spin-up and spin-down channels, respectively.

In order to map the configurations into the phase diagram, we first demonstrate that an A monolayer can be effectively described as a B monolayer plus an exchange field. Structurally, an A layer is formed by a B layer with MnTe intercalation. For a nonmagnetic B layer, the existence of inversion symmetry and $T$ ensures the spin degeneracy for the full Brillouin zone [Fig. 2(b)], while for an A layer the magnetization of Mn lifts the spin degeneracy [Fig. 2(c)]. The Mn-$3d^5$ states are located far away from the Fermi level (see Supplementary Section IV), implying that the main effect of Mn is to introduce a Zeeman exchange field to Bi$_2$Te$_3$. To prove this, we further project the band eigenstates onto $\sigma_z$, i.e. $\langle \varphi_n(\mathbf{k})|\sigma_z|\varphi_n(\mathbf{k})\rangle$, to distinguish the spin-up and spin-down channel. For the four bands around the Fermi level in the vicinity of $\Gamma$, the spin-up channels shifts upwards in energy compared with the spin-down channels, and the spin splittings for the valence band and the conduction band (0.15 eV and 0.16 eV, respectively) are almost the same. Hence, we conclude that for the four-band low-energy Hamiltonian an A layer can be well described by a B layer under an exchange field, with similar $g$ factor for different orbitals.

The topological properties of the 2D configurations, characterized by their Chern numbers ($C$) and spin Chern numbers ($C_s$), are listed in Table I. Figure 3 presents the 2D bulk, 1D edge state and Hall conductance of three representative topological phases, including $T$-preserved QSHE in BB (bilayer Bi$_2$Te$_3$), $T$-broken QSHE in AB, and QAHE in ABA. All of the three configurations have explicit band inversions between Bi-$p$ and Te-$p$ orbitals, but distinguished features for edge states. Specifically, for BB there is clear gapless Dirac cone protected by $T$, leading to a quantized spin Hall conductance in the bulk gap. In comparison, for AB the breaking of $T$ slightly gapped the Dirac cone, in consistent with the previous prediction [29]. As a result, the edge states consist of two counterpropagating channels with each almost spin polarized, resulting in a similar plateau of spin Hall conductance that exists in the clean limit but may not survive against disorder. Although the $Z_2$ index does not apply to systems without $T$, what makes AB topologically nontrivial is its nonzero spin Chern number, which remains valid when $T$ is broken and corresponds to a gapless projected spin spectrum given by $\langle\varphi_m(k_a)|\sigma_z|\varphi_n(k_a)\rangle$ [38]. For ABA configuration, the green squares in Fig. 3(h) show that there are two branches emerging from the valence band at $k_a = 0$ but only one back at $k_a = 2\pi$, indicating that the other branch connects the conduction band. Thus, the gapless edge state is contributed by only one spin channel as a single 1D chiral mode, leading to a quantized anomalous Hall conductance in the bulk gap. The band structure and edge state showing QAHE (ABBA and ABAB) as well as $T$-broken QSHE (BAB) for the other configurations is listed in Supplementary Section IV.

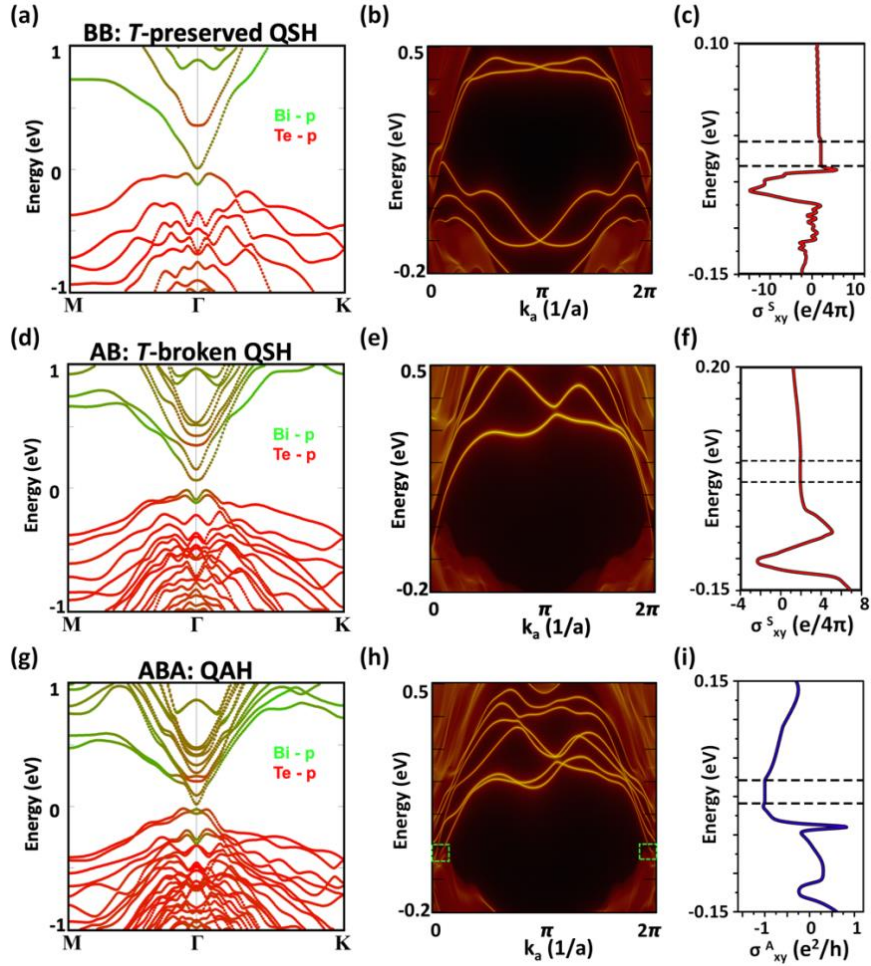

Fig. 3: (a-c) The band structure with projection onto different atomic orbitals (a), edge states (b) and quantized spin Hall conductance (c) of BB (bilayer $Bi_2Te_3$) configuration. (d-f) Same as (a-c) but for AB configuration. (g,h) Same as (a,b) but for ABA configuration. (i) quantized anomalous Hall conductance of ABA configuration.

*Design principles of quantum anomalous Hall effect*

Our DFT results suggest that the topological properties of the 2D systems fits the phase diagram quite well. This coincidence not only indicates that the main physics of the topological phase transitions is successfully captured by our simple model, but also provides us rational design principles of QAHE, as illustrated by Fig. 1(a). We put the configurations as discrete points into the phase diagram in the following way. Firstly, to begin with a specific configuration (e.g., AB), replacing a B layer by an A layer moves upward vertically (e.g., AA). Therefore, the evolution of BB-AB-AA [red vertical line in Fig. 1(a)] is to add exchange field gradually to a 2D *T*-preserved TI. The exchange field at first makes the band inversions of two spin channels nonequivalent, corresponding to *T*-broken QSHE, and then release the band inversion of one spin channel, turning the system to QAHE. Secondly, adding a nonmagnetic B layer moves rightward horizontally in the phase diagram. As shown by the red horizontal line in Fig. 1(a), the evolution of A-AB-BAB illustrates the phase transition between NI and *T*-broken QSHE by changing the film thickness.

We note that some conditions of our continuous model do not exist in the 2D systems we choose, such as the uniform exchange field applied to the film, and inversion symmetry that lacks in certain configurations. For example, ABAB and ABBA correspond to a same QAHE point in the phase diagram in terms of the thickness and total magnetization, but while ABBA has inversion symmetry, SIA effect exists in ABAB. Previous model Hamiltonian calculations reveal that the presence of SIA tends to release the band inversion in a Chern insulator by closing the band gap [26,34]. This is in consistent with our DFT-calculated results that both of ABAB and ABBA are Chern insulators with $C = 1$ but ABAB has a smaller band gap (see Table I). Another effect induced by Rashba SOC is that the 2D model Hamiltonian Eq. (2) is no longer block diagonal if SIA term is included, leading to hybridization between spin-up and spin-down channels. As a result, our DFT calculated spin Hall conductance for AB configuration is $1.94 \times \frac{e}{4\pi}$, slightly less than the exact quantization. However, we note that by choosing a proper "pseudo-spin" vector one can still block diagonal Eq. (2) with SIA term and then define a rigorous spin Chern number that is an integer for QSH insulator with *T*-broken [38].

The growth of new layered magnetic topological insulators such as MnBi$_4$Te$_7$ and MnBi$_6$Te$_{10}$ as well as the growth of VdW heterostructures by molecular beam epitaxy make it possible to realize different topological states by stacking different building blocks. Now the question is, how can we propose a candidate configuration, or unveil a "magic sequence" in a heterostructure, for a critically needed material functionality such as QAHE? The phase diagram [Fig. 1(a)] indeed helps to summarize the design principles. The region of QAHE indicates that in most cases, a sufficiently large magnetization is required. Therefore, a FM configuration should be the ground state, or metastable with a slightly higher total energy than the ground state. Due to the existence of Bi$_2$Te$_3$ buffer layer, a much smaller magnetic field than that of MnBi$_2$Te$_4$ thin films is required to trigger the QAHE in MnBi$_4$Te$_7$ or MnBi$_6$Te$_{10}$ films. Such magnetic manipulation can also be achieved by external magnetic field. By directly applying a magnetic field on AB configuration along the $c$ axis, we find that the phase transition between $T$-broken QSHE and QAHE also happens after a critical field $gM$ = 85 meV (see Supplementary Section V). Another possible strategy of such vertical regulation is to replace Mn$^{2+}$ (5/2 spin state) by another bivalent ion such as Eu$^{2+}$ with a higher 7/2 spin state, thus enhancing the spontaneous magnetization. In addition, with a fixed magnetization, QAHE could be realized by increasing the film thickness. This is because the band gap decreases with increasing film thickness, and so is the critical exchange field of topological phase transition. Such horizontal regulation could also be achieved continuously, and thus reach the QAHE region between A and AB, by electric field that applies different on-site energy to real-space separated orbitals. Consequently, through the exfoliation of MnBi$_4$Te$_7$ and MnBi$_6$Te$_{10}$ with superlattice-like stacking patterns, we are in principle able to reach a great area of the phase diagram discretely for the sought functionality, which calls for further experimental confirmation.

**Acknowledgments**

We thank Prof. Haizhou Lu, Ni Ni and Dr. Tong Zhou for helpful discussions. Work at SUSTech was supported by the NSFC under Grant No. 11874195.